# Identification of Common Trends in Political Speech in Social Media using Sentiment Analysis

Kostas Karpouzis [1[0000-0002-4615-6751]], Stavros Kaperonis [1[0000-0002-2130-6514]], and Yannis Skarpelos [1[0000-0002-8460-1676]]

[1] Department of Communication, Media and Culture,
Panteion University of Social and Political Sciences, Greece

**Abstract.** Social Media have been extensively used for commercial and political communication, besides their initial scope of providing an easy-to-use outlet to produce and consume user-generated content. Besides being a popular medium, Social Media have definitely changed the way we express ourselves or where we look for emerging news and commentary, especially during troubled times. In this paper, we examine a corpus assembled from the Twitter accounts of politicians in the United States and annotated with respect to their audience and the sentiment they convey with each post. Our purpose is to examine whether there are stylistic differences among representatives of different political ideologies, directed to different audiences or with dissimilar agendas. Our findings verify existing knowledge from conventional written communication and can be used to evaluate the quality and depth of political expression and dialogue, especially during the period leading to an election.

## 1 Introduction

The nature and content of political speech appears to have changed a lot since social media became prevalent. The impact of immediate communication offered by said platforms and the ability to spread a message quite quickly, besides allowing news outlets, politicians, and citizens to interact directly, has opened the door to adverse side-effects, such as hate speech (Mathew et al. 2019) and misinformation (Allcott et al. 2019). In this paper, we focus is on identifying whether the writing style and sentiment conveyed by social media posts related to politics differ depending on the target audience and the message each user is trying to convey. Even though social media posts are typically short, they have been shown to contain enough information to assist in identifying the age, cultural background or even the age of post originator (Zheng et al. 2006). The dataset used in this experiment was based on the "Political Social Media Posts - PSMP" (Figure Eight 2016) corpus, a part of the Data for Everyone project; it consists of 5000 tweets and Facebook posts with user and timestamp information, and annotations of audience (national or constituency), bias (neutral or partisan) and type of message (e.g., attack, information, mobilization, support, etc.)

## 2     Analysing Political Speech in Social Media

Feldman (2013) defines sentiment analysis as 'the task of finding the opinions of authors about specific entities'; for this reason, it's often termed 'opinion mining' in literature (Liu 2012), since the term 'sentiment' may be restrictive in that it points to concepts such as emotions and affect (Cowie et al. 2011a, Karpouzis and Yannakakis, 2016). In any case, before arriving to conclusions on high-level concepts, such as opinions and sentiments conveyed in text, computational approaches start by removing irrelevant information (such as "stop words" or location tags) and then proceed to look for word constructs with sentiment content. Initial methods (Taboada et al. 2011) relied on single words and looked them up in emotion lexicons, such as SentiWordNet (Baccianella et al. 2010). However, this approach is limited, since it misses out on language elements such as negation (e.g., 'not good' may be misinterpreted as a positive phrase, since 'not' is a stop word and 'good' has positive meaning) or language idiosyncrasies (e.g., phrases such as 'made my day' which is clearly positive but consists of words with neutral content). As a result, research methods quickly moved (Mohammad et al. 2013) to combinations of words, termed 'n-grams' (mostly bigrams, 3-grams) or longer sequences of words fed to neural networks (Wang et al. 2016). Utilizing neural networks, especially Deep Learning algorithms, provided a huge performance boost to sentiment analysis methods and allowed developers and users to analyse large corpora of text quickly, easily, and successfully (Godbole 2007). The downside of the computational methods based on machine learning is that they usually operate as a 'black box' (Zhang 2014) in the sense that we cannot identify which words, or other semantic or syntactic elements affected their estimate. As a result, these machine learning architectures are only as good (and as unbiased) as the data used to train them and any shortcomings will only surface during testing with different inputs: if, during training, the dataset is selected in such a manner that, for instance, positive texts come from sources of liberal speech and negative from conservative, there is a strong chance that a neural network might learn to classify all text from the first source as positive (and vice versa) regardless of their semantic content.

This issue is related to Julia Flanders commentary on computational methods in (Flanders 2016), where she criticizes the 'rhetoric of abundance' when she discusses digital resource development and mentions Unsworth (2002) by stating that humanities computing is mostly about modelling knowledge and even modelling the modelling process. In essence, she describes a 'tug of war' between this understanding and the incessant utilization of computational methods to provide estimations or identify relations where they might not really exist. Admittedly, we now have the opportunity to quickly process vast amounts of diverse and multimodal data, from digitized paint-ings and 3D artefacts to selfies (cf. the SelfieCity[1] project, Tifentale and Manovich 2015) and social media posts; however, this does not necessarily mean that all findings of computational methods will make sense or be useful (cf. Manovich 2017 for a preference of the left cheek identified in entries of the SelfieCity image database) or even be fair and unbiased (Kiritchenko & Mohammad 2018). What would be far more useful and beneficial for both Digital Humanities (DH) and Computer Science would be the 'gradual shift' towards a digital transformation of methods, described by Jukka Tyrkkö (2007); in his work, Tyrkkö describes DH as the intersection of qualitative perspectives (where the questions come from) and quantitative, computational methods that (hopefully) provide the answers. In the context of this paper, the aim of the analysis was not to draw conclusions about how the opinions of social media users are expressed online, but to investigate whether posts from different perspectives, targeted towards different audiences and with a different purpose may share a common style or emotional content; this is consistent with Tyrkkö's opinion about not using large scale computational processing to make statistical predictions about the general population, but to categorize content and authors based on style similarity or similarity of topic.

The latter constitutes a valuable lesson for social media users: we are creating publicly available content every time we post a selfie, a status update or a tweet, and this content can be viewed, recorded, or processed not only by the service we're using, but also by search engines or other users, unknown to or socially disconnected from us. The Findability concept discussed by Jakobsson (2021) may not initially be defined for social media data, but is still extremely relevant here, since '[d]ata can be found by anyone, at any time, and from anywhere'. Research (Sakariassen and Meijer 2021) has shown that social media users sometimes feel more empowered to express themselves online, because of either the relative anonymity offered by the medium or the physical distance that separates people interacting via direct messages or chats. In any case, social media make it easier for other people to retrieve our posts, by providing Application Programming Interfaces (APIs) for easy bulk access, enhancing the Accessibility and Interoperability (Jakobsson 2021) of the content in the process. Regarding the Re-usability of the content, there are legal and social issues related to identifying users, processing their posts and using it for other (e.g., commercial) purposes (Ahmed et al. 2017), but for the scope of this paper, the dataset contains no identifying information which could help someone trace each post back to its originator.

---

[1] Selfiecity, Investigating the style of self-portraits (selfies) in five cities across the world, retrieved from https://selfiecity.net/ on 13 September 2022.

## 3  Analysis and results

The PSMP dataset is a subset of the "Data for Everyone Library"[2] and provides text of 5000 messages from politicians' (US Senators and other American politicians) social media accounts, along with human annotations about the purpose, partisanship, and audience of the messages. The data are broken down into audience (*national* or the tweeter's *constituency*), bias (*neutral/bipartisan* or *biased/partisan*), and tagged as the actual substance of the message itself; options include:

- attack: the message attacks another politician
- constituency: the message discusses the politician's constituency
- information: an informational message about news in government or the wider U.S.
- media: a message about interaction with the media
- mobilization: a message intended to mobilize supporters
- other: a catch-all category for messages that don't fit into the other
- personal: a personal message, usually expressing sympathy, support or condolences, or other personal opinions
- policy: a message about political policy, and
- support: a message of political support

**Fig. 1.** The PSMP download page with a selection of the content and its assorted metadata

---

[2] Data for Everyone, Pre-labeled datasets, retrieved from https://appen.com/pre-labeled-datasets/ on September 13, 2022.

The first step towards modelling the data was to keep only posts that contained enough linguistic content to process; that meant that posts containing only links to webpages had to be removed. Posts consisting solely of hashtags were also removed, since they are sometimes difficult to annotate because of each user personal style (Rauschnabel et al., 2019). To this end, OpenRefine[3] was used, utilizing its ability to filter posts using regular expressions (Verborgh and De Wilde 2013).

For the analysis part, the spaCy Python library (Honnibal and Montani 2017) was used to estimate the sentiment conveyed in each of the remaining posts; besides this value, spaCy returns the degree of certainty for each estimate and the words with high sentiment content in each message. The latter annotations were used to discard posts with ambiguous sentiment (subjectivity < 0.6) and to check whether posts with similar annotations used the same vocabulary.

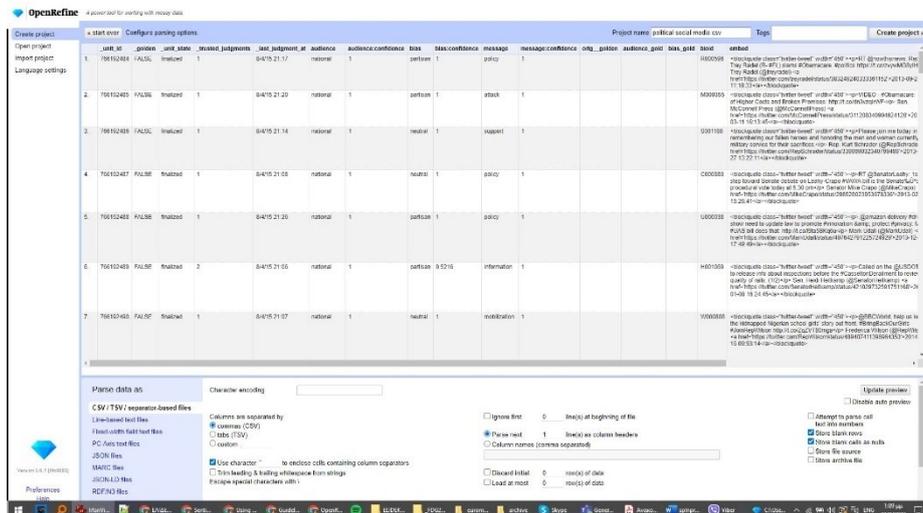

**Fig. 2.** OpenRefine user interface showing the PSMP dataset and metadata

### 3.1 Results and discussion

What was evident from analysing the PSMP dataset was that the sentiment conveyed by each post depended a lot on the target audience, the bias, and the nature of the message, while posts annotated as *'informative'* were rarely emotional. More specifically, only 13.4% of informative posts were emotionally rich, while strong sentiment was conveyed by more than 92% of posts annotated as *'attack'* (negative), *'support'* or *'mobilization'* (both positive). While the latter finding was expected, since a message used to attack someone has to include negative words about them or calls to violence (which, again, have negative connotations), it was refreshing to see that informative posts were, indeed, neutral and did not follow the recent trend of introducing a narrative structure to news stories; Machill (2007) argues that many news outlets like to intertwine concepts from narration (for instance, heroes vs villains, objects of desire, a call to action, etc.) so as to facilitate easier and deeper rapport from viewers, sacrificing their objectivity to a large extent. Perhaps this trend can be attributed to the limitation imposed by Twitter on the length of posts and the reluctance of Facebook users to engage with longer posts, which has forced news content creators to refrain from such posts and keep narrative-rich stories for website and blog posts.

---

[3] OpenRefine, A free, open source, power tool for working with messy data, retrieved from http://www.openrefine.org on September 13, 2022.

Another interesting finding had to do with the audience of each post, with more posts characterised as being targeted towards constituency being emotionally rich (78.3%, as opposed to 65.2% of posts with a national audience). This is consistent with research by political scientists (e.g., Stier 2018) about the differences in communication style when politicians and policy makers address different audiences, but it may also reflect cultural differences thought to exist in the targeted viewers; this became apparent with the rise of populist politics in Europe and the USA, who tend to use specific language traits in their communication (Block and Negrine 2017), especially when addressing people from working-class or lower-income districts.

The final research question had to do with how rich the vocabulary used in emotional posts was. To answer this, we used spaCy to extract emotionally rich words from each tweet or Facebook post, using only messages with high sentiment value; those words were aggregated, taking two categories into account, namely *'neutral'* and *'partisan'* as bias tags. This 'bag-of-words' approach (Wang et al. 2014) is useful when we are not that much interested in each post and its properties as a separate entity, but we want to identify possible trends or common characteristics of a group of posts. In this case, posts tagged as *'partisan'* utilised a narrower range of emotional words (0.3 discrete words per post) than *'neutral'* (0.7 discrete words per post). Wang and Liu (2018) argue that populist politicians rarely make use of a rich vocabulary, since this may sometimes be miscomprehended by partisan audience, and prefer to use fewer and emotionally strong words.

## 4    Conclusion

In this paper, we investigated shared trends in political speech in social media, based on target audience, sentiment content and purpose of each post. The computational part of this was carried out using open access tools (OpenRefine and Python), providing quantitative estimates (counts and means) for different categories, as well as individual posts. We then attempted to make sense of the values returned by the computational methods, based on literature related to political speech and expression in social media. As mentioned by Tyrkkö (2007), the appeal of this discourse between quantitative processing and qualitative examination of those results comes from the ability of the former to spot things that humans are unable to and from making "humanities research more verifiable, replicable and reliable". Using readily available machine learning libraries and ubiquitous, cloud-based computing infrastructures we can now process the vast amount of user-generated cultural content and make sense of it or, sometime, completely revisit our understanding of the world.